\documentclass[%
 reprint,
superscriptaddress,
 amsmath,amssymb,
 aps,
floatfix,
showkeys
]{revtex4-2}

\usepackage{graphicx}
\usepackage{dcolumn}
\usepackage{bm}
\usepackage[utf8]{inputenc}
\usepackage{tabularx}
\usepackage[dvipsnames]{xcolor}
\usepackage{units}
\usepackage[euler]{textgreek}
\usepackage{upgreek}
\usepackage{derivative}
\usepackage{amsmath}
\usepackage{physics}
\usepackage{amssymb}
\usepackage[colorlinks=true,linkcolor=blue!50!black,urlcolor=blue!50!black,citecolor=blue!50!black]{hyperref}
\usepackage{afterpage}
\usepackage{float}

\begin{document}

\preprint{APS}

\newcommand{\mpi}{\affiliation{Max-Planck-Institut f\"ur Physik, 80805 M\"unchen, Germany}}
\newcommand{\coimbra}{\affiliation{Also at: LIBPhys, Departamento de Fisica, Universidade de Coimbra, P3004 516 Coimbra, Portugal}}
\newcommand{\hephy}{\affiliation{Institut f\"ur Hochenergiephysik der \"Osterreichischen Akademie der Wissenschaften, 1050 Wien, Austria}}
\newcommand{\ati}{\affiliation{Atominstitut, Technische Universit\"at Wien, 1020 Wien, Austria}}
\newcommand{\tum}{\affiliation{Physik-Department, TUM School of Natural Sciences, Technische Universit\"at M\"unchen, D-85747 Garching, Germany}}
\newcommand{\tuebingen}{\affiliation{Eberhard-Karls-Universit\"at T\"ubingen, 72076 T\"ubingen, Germany}} 
\newcommand{\bratislava}{\affiliation{Comenius University, Faculty of Mathematics, Physics and Informatics, 84248 Bratislava, Slovakia}}

\newcommand{\oxford}{\affiliation{Department of Physics, University of Oxford, Oxford OX1 3RH, United Kingdom}}
\newcommand{\wmi}{\affiliation{Also at: Walther-Mei\ss ner-Institut f\"ur Tieftemperaturforschung, 85748 Garching, Germany}}
\newcommand{\lngs}{\affiliation{INFN, Laboratori Nazionali del Gran Sasso, 67010 Assergi, Italy}}
\newcommand{\gssi}{\affiliation{Also at: Gran Sasso Science Institute, 67100, L'Aquila, Italy}}
\newcommand{\cassino}{\affiliation{Also at: Dipartimento di Ingegneria Civile e Meccanica, Università degli Studi di Cassino e del Lazio Meridionale, 03043 Cassino, Italy}}
\newcommand{\usp}{\affiliation{Also at: Instituto de F\'{i}sica, Universidade de S$\tilde{a}$o Paulo, S$\tilde{a}$o Paulo 05508-090, Brazil}}
\newcommand{\umb}{\affiliation{Also at: Dipartimento di Fisica, Universit\`a di Milano Bicocca, Milano, 20126, Italy}}

\mpi
\lngs
\tum
\hephy
\ati
\tuebingen
\oxford
\bratislava

\coimbra
\gssi
\wmi
\cassino
\usp
\umb

\author{G.~Angloher}
  \mpi

\author{S.~Banik}
  \hephy
  \ati

\author{G.~Benato}
  \lngs
  \gssi

\author{A.~Bento}
  \mpi
  \coimbra 

\author{A.~Bertolini}
  \mpi

\author{R.~Breier}
  \bratislava

\author{C.~Bucci}
  \lngs 

\author{J.~Burkhart}
  \hephy

\author{L.~Canonica}
  \mpi 

\author{A.~D'Addabbo}
  \lngs

\author{S.~Di~Lorenzo}
  \mpi

\author{L.~Einfalt}
  \hephy
  \ati
  
\author{A.~Erb}
  \tum
  \wmi
  
\author{F.~v.~Feilitzsch}
  \tum  
  
 \author{S.~Fichtinger}
  \hephy
 
\author{D.~Fuchs}
    \email[Corresponding author: ]{dominik.fuchs@mpp.mpg.de}
  \mpi  

\author{A.~Garai}
  \mpi 
  
 \author{V.M.~Ghete}
  \hephy 

\author{P.~Gorla}
  \lngs 

\author{P.V.~Guillaumon}
  \mpi
  \usp

 \author{S.~Gupta}
  \hephy 

\author{D.~Hauff}
  \mpi 

\author{M.~Ješkovsk\'y}
  \bratislava

\author{J.~Jochum}
  \tuebingen 

\author{M.~Kaznacheeva}
  \tum

\author{A.~Kinast}
  \tum
  
\author{H.~Kluck}
  \hephy

\author{H.~Kraus}
  \oxford

\author{S.~Kuckuk}
  \tuebingen 

\author{A.~Langenk\"amper}
  \mpi

\author{M.~Mancuso}
  \mpi
 
 \author{L.~Marini}
  \lngs

\author{B.~Mauri}
    \mpi

\author{L.~Meyer}
  \tuebingen 
  
\author{V.~Mokina}
  \hephy

\author{M.~Olmi}
  \lngs
  
\author{T.~Ortmann}
  \tum

\author{C.~Pagliarone}
  \lngs 
  \cassino

\author{L.~Pattavina}
  \lngs
  \umb

\author{F.~Petricca}
  \mpi 

\author{W.~Potzel}
  \tum 

\author{P.~Povinec}
  \bratislava


\author{F.~Pr\"obst}
  \mpi

\author{F.~Pucci}
  \mpi 
  \tum
  
\author{F.~Reindl}
  \hephy
  \ati

\author{J.~Rothe}
  \tum
  
\author{K.~Sch\"affner}
  \mpi

\author{J.~Schieck}
  \hephy
  \ati 

\author{S.~Sch\"onert}
  \tum 
  
\author{C.~Schwertner}
  \hephy
  \ati

\author{M.~Stahlberg}
  \mpi

\author{L.~Stodolsky}
  \mpi 

\author{C.~Strandhagen}
  \tuebingen

\author{R.~Strauss}
  \tum

\author{I.~Usherov}
  \tuebingen 

\author{F.~Wagner}
  \hephy

\author{V.~Wagner}
    \tum

\author{V.~Zema}
  \mpi

\collaboration{CRESST Collaboration}
\noaffiliation

\title{First observation of single photons in a CRESST detector and new dark matter exclusion limits}

\begin{abstract}
The main goal of the CRESST-III experiment is the direct detection of dark matter particles via their scattering off target nuclei in cryogenic detectors. In this work we present the results of a Silicon-On-Sapphire (SOS) detector with a mass of 0.6$\,$g and an energy threshold of (6.7$\, \pm \,$0.2)$\,$eV with a baseline energy resolution of (1.0$\, \pm \,$0.2)$\,$eV. This allowed for a calibration via the detection of single luminescence photons in the eV-range, which could be observed in CRESST for the first time. We present new exclusion limits on the spin-independent and spin-dependent dark matter-nucleon cross section that extend to dark matter particle masses of less than 100$\,$MeV/c$^{2}$.
\end{abstract}

\maketitle



\section{Introduction}
The search for dark matter (DM) is one of the biggest challenges of modern-day physics.
Convincing evidence for the existence of this non-luminous matter has been found
on various scales of the Universe over the course of the last century. However, despite
many experimental efforts, the nature of DM is still to be unveiled. An appealing solution to this problem is the introduction of one or more new unknown particles \cite{ARBEY2021103865}.
The CRESST-III (Cryogenic Rare Event Search with Superconducting Thermometers)
experiment's main goal is the direct detection of DM particles via their scattering off target nuclei in cryogenic detectors. The detectors are equipped with transition edge sensors (TESs), operated at around 15 mK. These detectors reach sensitivities down to very low energy depositions ($\leq \,$100$\,$eV), allowing for the search of DM particles with sub-GeV/c$^{2}$ masses. This makes CRESST one of the leading experiments in low-mass DM searches. 
Another advantage of this technology is the possibility to use different target materials. In recent measurement campaigns of CRESST-III various materials have been used as targets: CaWO$_{4}$, Si, LiAlO$_{2}$, Al$_{2}$O$_{3}$ (sapphire) and Silicon-On-Sapphire (SOS). Currently, the strongest limits on spin-independent interactions with sub-GeV/c$^{2}$ DM under standard assumptions were obtained with a CaWO$_{4}$ detector \cite{PhysRevD.100.102002} and a Si detector \cite{PhysRevD.107.122003}. The strongest limits on spin-dependent interactions of DM with protons and neutrons were obtained with a LiAlO$_{2}$ detector \cite{PhysRevD.106.092008}.

In this paper we present new results obtained with a SOS detector, operated between November 2020 and August 2021 with a total exposure of 0.138$\,$kg$\,$d. Due to the low threshold of 6.7$\,$eV and the baseline energy resolution of 1.0$\,$eV in this detector, we are able to measure and resolve single photons in a cryogenic solid state detector based on the TES technology for the first time within CRESST. We observe vacuum ultra violet (VUV) luminescence photons from a nearby Al$_2$O$_{3}$ crystal, which we use to fine-tune our low energy calibration. The low energy threshold of this detector further allows to extend our sensitivity to DM particle masses of less than 100$\,$MeV/c$^{2}$.

First, we describe the experimental setup and the detector design in Sec. \ref{Setup}. The data analysis, in particular our novel way of energy calibration, based on the detection of 7.6$\,$eV luminescence photons, is described in Sec. \ref{Analysis}. Then we present the new exclusion limits on spin-independent and spin-dependent DM interactions in Sec. \ref{DMResults} and conclude the discussion in Sec. \ref{Conclusion}.

\section{Experimental setup and detector design} \label{Setup}
\subsection{Experimental setup}
Since the expected rate of a DM signal is extremely low, a low-background environment is necessary. The CRESST experiment is therefore located in the underground facility of the Laboratori Nazionali del Gran Sasso (LNGS) in Italy. The rock overburden of about 1400$\,$m (corresponding to 3800$\,$m.w.e.) reduces the flux of cosmic muons to (3.41$\, \pm \,$0.01)$\, \cdot 10^{-4} \,$m$^{-2} \,$s$^{-1}$ \cite{Bellini_2012}. Remaining muons are tagged by a muon veto system with a geometric coverage of 98.7$\,$\% \cite{ANGLOHER2009270}. The experiment is additionally shielded by concentric layers of various shielding materials. Environmental neutrons are moderated by the outermost layer of polyethylene. 
Layers of lead and copper shield against radioactive backgrounds. Finally the detectors are surrounded by another layer of polyethylene, protecting against neutrons originating from the lead or copper shields. 
The extremely low temperatures that are required to operate the CRESST detectors ($\mathcal{O}$(mK)) are achieved by the use of a commercial $^{3}$He/$^{4}$He-dilution refrigerator. For a more detailed description of the experimental setup see \cite{ANGLOHER2009270}.

\subsection{Detector Design} \label{DetDesign}
\begin{figure}[h]
\centering
\includegraphics[width=0.4\textwidth]{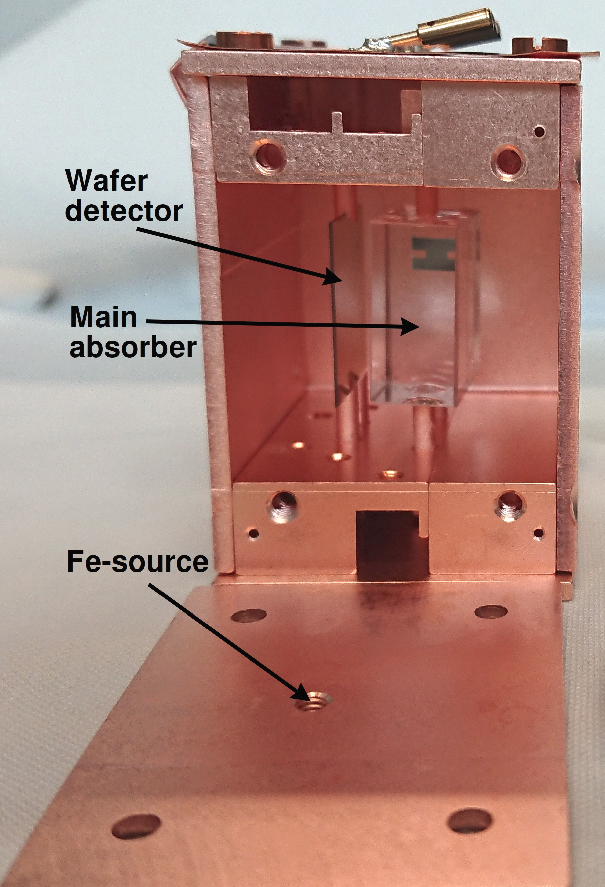}
\caption{\label{fig:SappMod} Picture of the sapphire detector module. Shown are the main absorber, consisting of monocrystalline sapphire, and the SOS wafer detector. Both detectors are equipped with a W-TES. The crystals are held by copper sticks in a bare copper housing. The $^{55}$Fe calibration source is mounted through a small hole in one of the side walls of the module.}
\end{figure}

The standard design of a CRESST detector module consists of a main absorber and a wafer detector. This work concentrates on the analysis and results of the wafer detector of a sapphire detector module. The full module is shown in Fig. \ref{fig:SappMod}. The main absorber is an (20$\,$x$\,$20$\,$x$\,$10)$\,$mm$^{3}$ Al$_{2}$O$_{3}$ crystal, equipped with a tungsten transition edge sensor (W-TES). The wafer detector, facing the main absorber, is a (20$\,$x$\,$20$\,$x$\,$0.4)$\,$mm$^3$ SOS crystal (silicon layer of $\sim \,$1$\,$\textmu m on an Al$_{2}$O$_{3}$ crystal), likewise equipped with a W-TES. Both crystals are held by copper sticks and are kept in a bare copper housing. In both detectors, the W-TES film is operated at a stable temperature around $\sim \,$15$\,$mK, corresponding to a point in the transition between the normal conducting and super conducting phase, where its electric resistance depends linearly on its temperature. Signals are converted into a voltage pulse by SQUIDs and are subsequently read out and saved by the data aquisition system \cite{Angloher2012}. The detectors are sensitive to temperature changes of $\mathcal{O}$(\textmu K). The sensors are stabilized at their operating temperature via heating resistors on the crystals. These resistors are used to periodically inject electric pulses (in the following called heater pulses) at different energies into the system, covering the dynamic range of the TES. This allows to determine the detector resolution over the full range of the TES. In addition to the stabilization of the detectors, these heater pulses are also used to monitor the time dependence of the detector response to energy depositions, which is used in the calibration. A low-activity $^{55}$Fe source, used for energy calibration of the detectors, is mounted through a small hole in the side wall of the module. The source is covered by a layer of glue and coated with gold to shield against Auger electrons and to prevent scintillation light originating from the glue to reach the detector.

\section{Data analysis} \label{Analysis}
\subsection{Data processing}
We use a continuous data acquisition with which the full, dead-time free output of the detector is recorded and saved to disk \cite{ediss23762}. 
The triggering of the data is done offline with a software using an optimum filter \cite{GattiManfredi1986} and can be iteratively repeated as the analysis is refined. Furthermore, the full stream of detector output is available, which contains not only the particle and heater pulses, but also a large fraction of noise. This is used to perform simulations by superimposing artificial pulses onto the stream, which are necessary in order to estimate the survival probability of events (see Sec. \ref{SpecAndEff}). For this work we used data recorded between November 2020 and August 2021 with the detector described in Sec. \ref{DetDesign}. Before cuts, these data amount to a gross exposure of 0.138$\,$kg$\,$d.

In a first step, we maximize the signal-to-noise ratio of the recorded data with an optimum filter, which is individually created, based on the specific signal pulse shape and noise characteristics of each detector. Due to the design of this filter, the original pulse height of the signals is preserved, such that the filter amplitude is a direct estimator for the signal amplitude.
We then define the trigger threshold by setting the accepted number of noise triggers to 1$\,$kg$^{-1}$$\,$d$^{-1}$ and select all pulses with amplitudes above this threshold. The threshold determination is based on an analytical model of the distribution of amplitudes of a large ensemble of filtered noise samples, which allows to calculate the noise trigger rate as a function of the threshold. This method is described in detail in \cite{MANCUSO2019492}. The detector presented in this work has an extremely low threshold at only (6.7$\, \pm \,$0.2)$\,$eV.

The amplitudes of the pulses carry the information of the energy deposited in the detector. All triggered pulses are subject to a list of selection criteria. The data selection criteria are designed to keep only pulses for which the amplitude can be properly reconstructed. A detailed list of the applied selection criteria can be found in \cite{fuchs2023}.

\subsection{Calibration}
The wafer detector is optimized for very low energies. Therefore, the direct hits of the events coming from the K$_{\alpha}$ and K$_{\beta}$ X-ray lines of the $^{55}$Fe decay at 5.89$\,$keV and 6.49$\,$keV \cite{international2007iaea} are energetic enough to drive the TES out of its transition into the normal conducting phase, leading to highly saturated pulses. The amplitude of these pulses cannot be reconstructed with an optimum filter. Instead we perform a truncated standard event fit which uses the known pulse shape information only in the linear regime of the TES, excluding all pulse samples above the corresponding truncation limit. 
This method is well established and was used in past CRESST analyses \cite{Angloher2014, Angloher2016}.
Here we use this method for a first calibration of the wafer detector.

\begin{figure}[!h]
\centering
\includegraphics[width=0.5\textwidth]{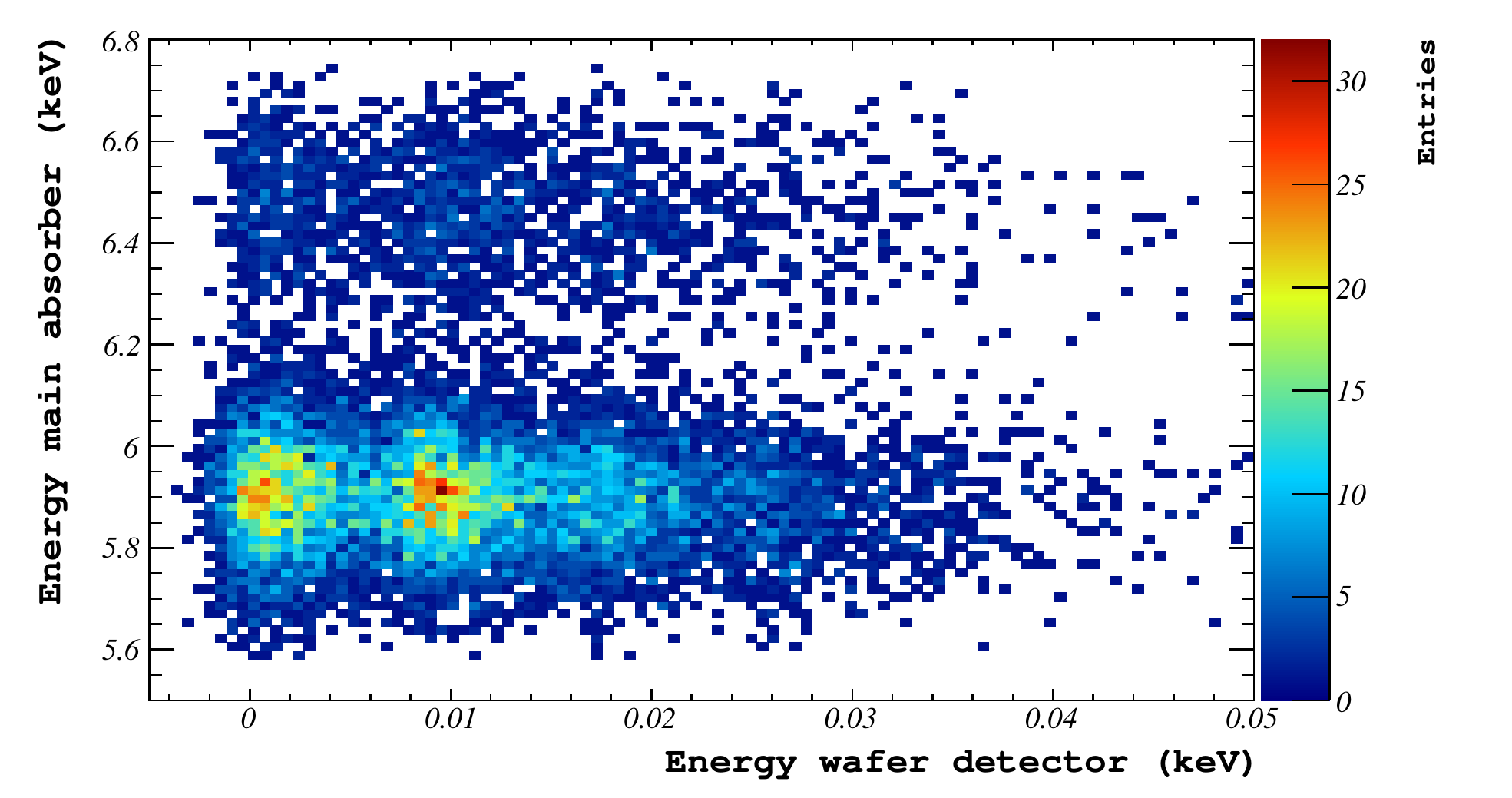}
\caption{\label{fig:Peaks2D} 2D histogram of the energies in both detectors of events coincident with hits of X-rays from the $^{55}$Fe calibration source at 5.89$\,$keV and 6.49$\,$keV (y-axis) in the main absorber. The energies in the wafer detector (x-axis) show several separate equidistant populations at very low energies.}
\end{figure}

\paragraph{Luminescence of the main absorber:}
we observe a coincident signal between the main absorber and the wafer detector, which is mostly visible in the case of $^{55}$Fe X-rays hitting the main absorber. Figure \ref{fig:Peaks2D} shows a 2D histogram of these signals in both detectors. The energy scale of the wafer detector in this plot is the one of the calibration based on the truncated fit, as described above. The energy depositions in the wafer detector show several equidistant populations, with the first population being centered around zero and the second one following at around 9$\,$eV.

Ultra pure sapphire crystals emit VUV luminescence at 7.6$\,$eV via the radiative decay of excitons \cite{RUNCIMAN1968537, PhysRevB.60.502}. This luminescence can be actively excited by external radiation. The high initial energy deposition in the main absorber of about 6$\,$keV can lead to a multiplication of the electronic excitations, which in turn leads to the creation of several luminescence photons, each with an energy of 7.6 eV \cite{PhysRevB.60.502}. The timescale of the VUV emission is of the order of ns \cite{PhysRevB.60.502, LUSHCHIK2000232, 10.1117/12.639064}, while the pulse risetime for the wafer detector is of the order of 100$\,$\textmu s. Therefore, multiple luminescence photons from one X-ray event cannot be resolved in time and their energies will add up. 

The probability of a single luminescence photon to deposit energy in the wafer detector is mostly given by the solid angle of the wafer detector with respect to the position of the emission in the main absorber. The probability of
measuring $n \in \lbrace0,1,2,3,4,...\rbrace$ photons in the detector simultaneously is then described by a binomial distribution, which explains the observed decreasing intensity between the equally spaced populations in Fig. \ref{fig:Peaks2D}. We therefore interpret these populations as the energy depositions of single photons of 7.6$\,$eV (second population) and multiples of such (following populations). The first population (centered around zero) shows the case of no luminescence photons reaching the wafer detector. Indeed, we are able to identify these events as empty noise traces and remove the first population with our selection criteria.

\begin{figure}[!h]
\centering
\includegraphics[width=0.5\textwidth]{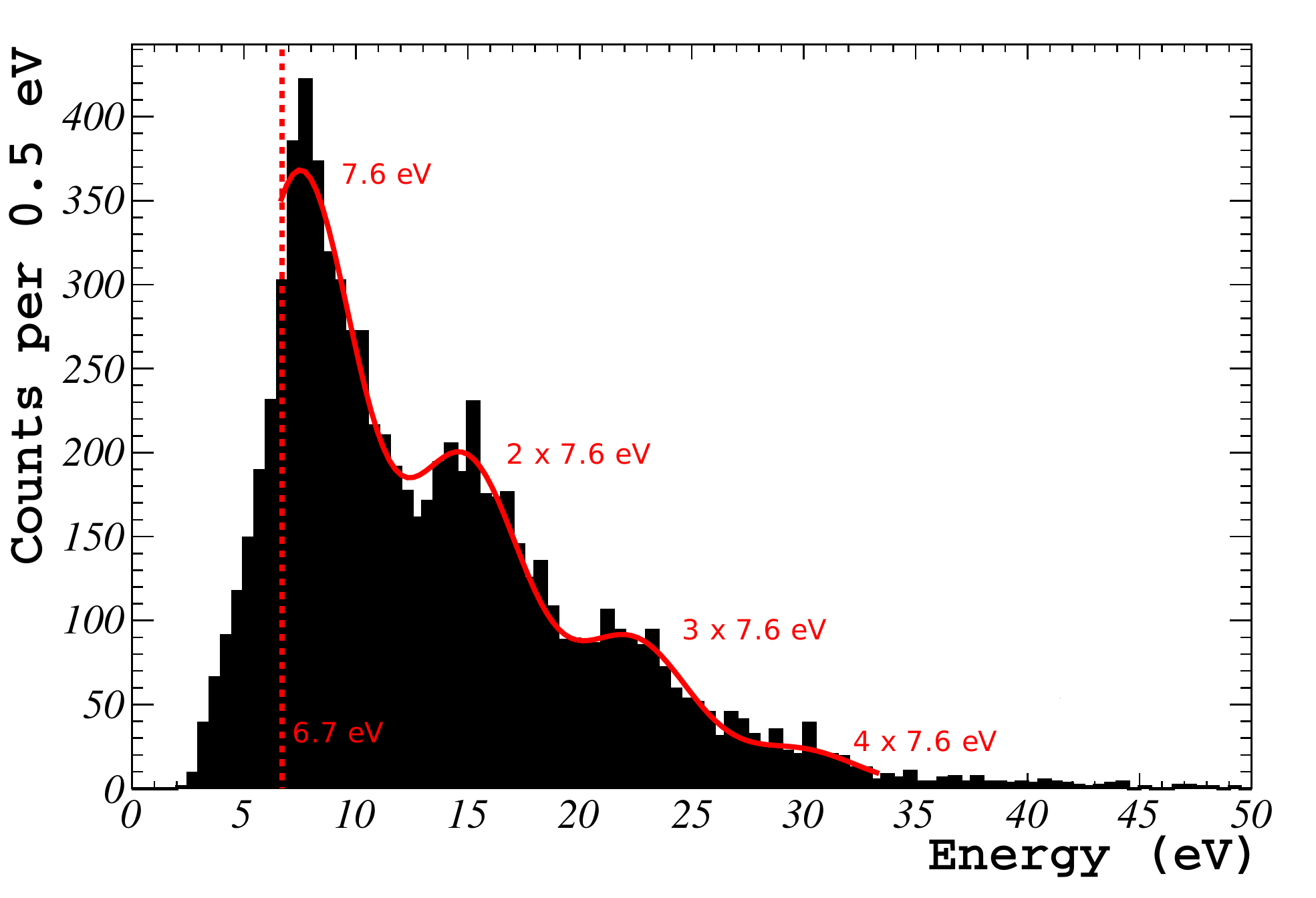}
\caption{\label{fig:Peaks1D} Energy spectrum of luminescence events in the wafer detector in coincidence with $^{55}$Fe events in the main absorber (black) together with four fitted Gaussian distributions with equal distance and resolutions (red). The red dashed line depicts the energy threshold after the fine-tuning of the calibration.}
\end{figure}

The presence of these peaks is an excellent opportunity to perform a fine-tuning of the calibration at these low energies. For this we fit four Gaussian distributions with equal distance and constant resolution to a histogram of the reconstructed amplitudes of the events forming the luminescence peaks. The fit is restricted to amplitudes above the trigger threshold of the wafer detector in order to ensure that they are correctly reconstructed. The resulting conversion factor from reconstructed amplitudes to units of energy decreases with respect to the original calibration factor based on the fit of the saturated pulses of the direct hits of $^{55}$Fe events in the wafer detector. This shifts the threshold in units of energy from 8.0$\,$eV (original calibration) to 6.7$\,$eV (fine-tuned calibration). Figure \ref{fig:Peaks1D} shows a histogram of the luminescence events after the fine-tuning of the calibration together with four fitted Gaussian distributions.

\subsection{Final spectrum and efficiency} \label{SpecAndEff}
\begin{figure}[h]
\centering
\includegraphics[width=0.5\textwidth]{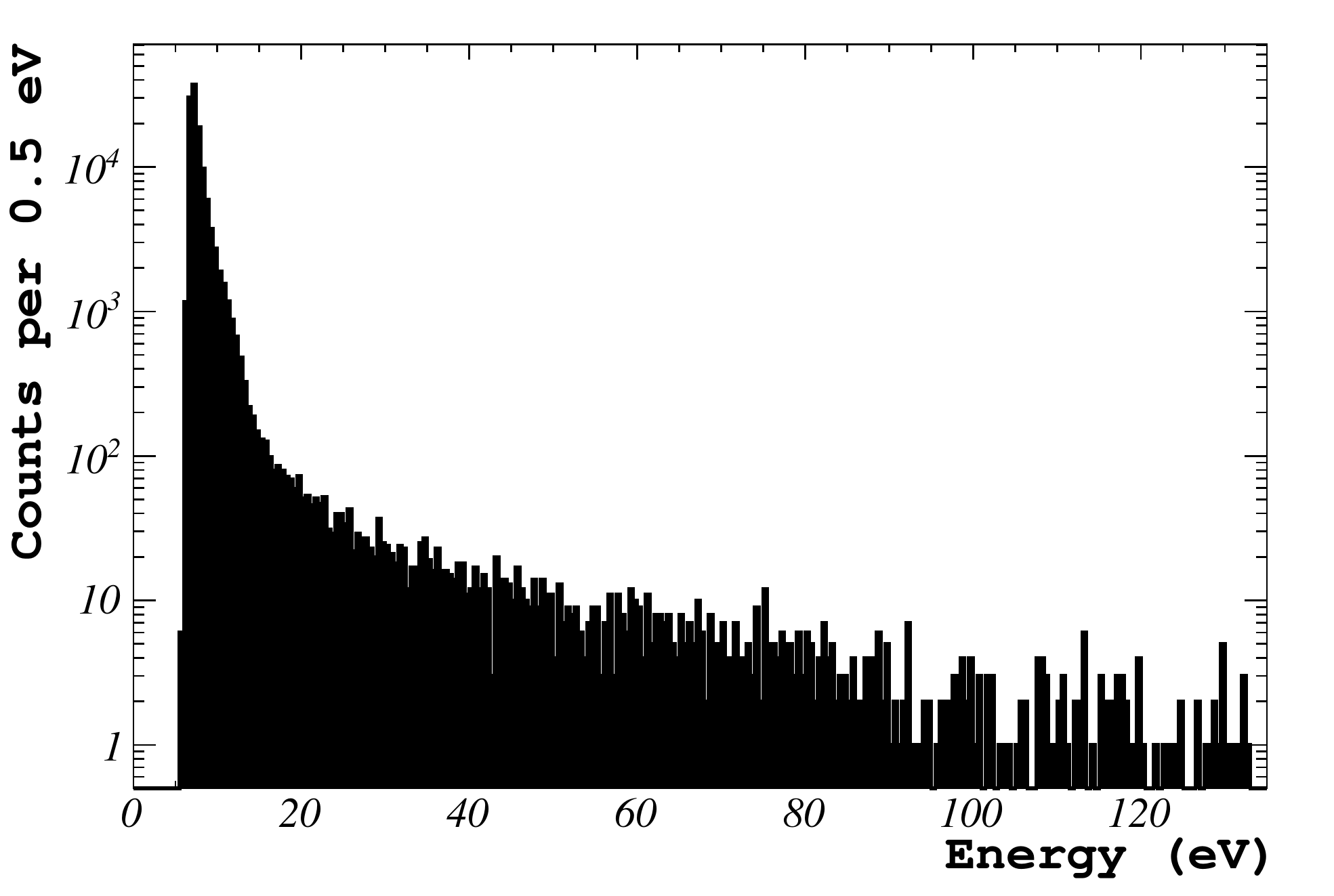}
\caption{\label{fig:Spec} Final energy spectrum used for the DM analysis, after applying all selection criteria, with a total exposure of 0.138$\,$kg$\,$d. The steeply rising event rate towards lower energies is known as the low energy excess (LEE).}
\end{figure}

In the final calibrated energy spectrum used for the DM analysis, all events that are clearly in coincidence with an $^{55}$Fe event in the main absorber (the luminescence peaks) are removed by our selection criteria. The final energy spectrum, measured with a total exposure of 0.138$\,$kg$\,$d and after applying all selection criteria, is shown in Fig. \ref{fig:Spec}. The most striking feature of the spectrum is the strong rise of the event rate towards lower energies. The rate of these low energy events strongly exceeds the expected background rate and is therefore often referred to as the low energy excess (LEE). An extensive summary of the observed LEE in CRESST and other experiments can be found in the report of the first EXCESS Workshop \cite{10.21468/SciPostPhysProc.9.001} and a deeper investigation of the LEE in CRESST in \cite{10.21468/SciPostPhysProc.12.013}.

\begin{figure}[h]
\centering
\includegraphics[width=0.5\textwidth]{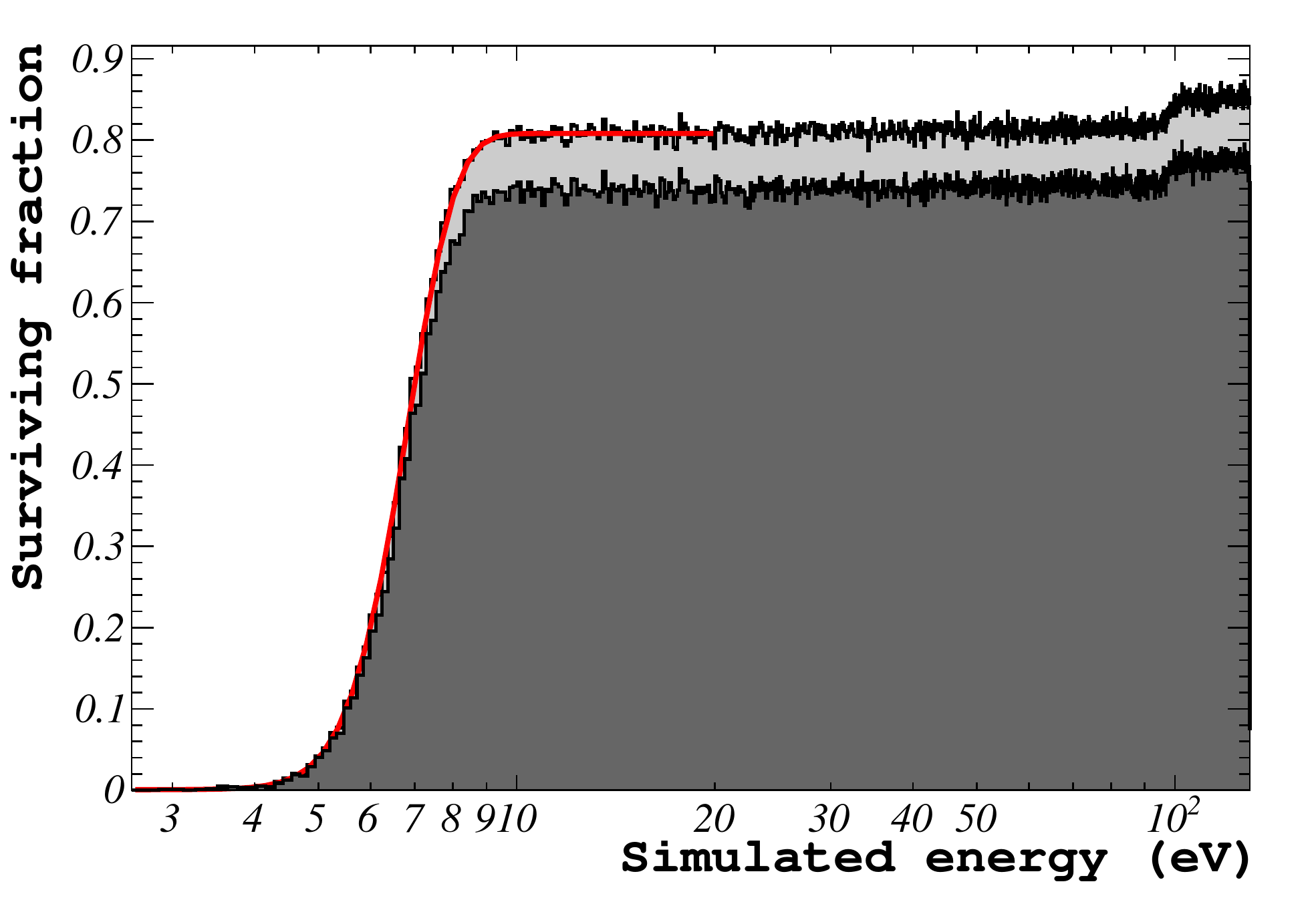}
\caption{\label{fig:Eff} Energy dependent surviving fraction of simulated events after triggering (light grey) and after applying all selection criteria (dark grey). The red line shows a fit of an error function to the trigger efficiency, verifying the energy threshold at 6.7$\,$eV and a baseline resolution of 1.0$\,$eV.}
\end{figure}

The probability of a valid event to survive all steps of the analysis is obtained by performing a simulation. We superimpose $\sim \,$3$\, \times \,$10$^{6}$ artificial pulses onto the continuous stream of real data at random times to obtain a simulated data stream with known energies per event. Their energies are uniformly distributed over the linear regime of the detector between 0 and 130$\,$eV. The simulated data are then processed by the same analysis chain as the real data. Due to the randomness in the time distribution, some of the pulses are superimposed onto different types of artifacts on the stream with a proportion that is representative of the real data. Since the amplitudes and timestamps of the simulated events are known, the surviving fraction of events after triggering and after applying the selection criteria can be extracted as a function of their energy. This energy dependent surviving fraction of events can then be used as an estimator for the survival probability of real particle events. Figure \ref{fig:Eff} shows the energy dependent fraction of surviving events after triggering (light gray) and after applying all selection criteria (dark gray).

The constant trigger survival probability of $\sim \,$81$\,$\% above 10$\,$eV is caused by dead time of the trigger, mostly due to the presence of heater pulses that are injected into the detectors for stabilisation and calibration purposes. The shoulder in the efficiency curves at around 0.1$\,$keV emerges from a filter effect. The filtered output of a pulse contains small side maxima, symmetrically distributed around the pulse \cite{ediss23762}. The height of these maxima depends on the amplitude of the originally filtered pulse. The amplitude of the side maxima of the largest heater pulses corresponds to an energy of about 0.1$\,$keV in this detector. After the filter process, particle events with energies below this value that appear close enough to such a heater pulse in time (before or after) can be hidden by one of these side maxima. Effectively this reduces the trigger probability for particle events with energies below
0.1$\,$keV, leading to the shoulder in the efficiency curve.

Figure \ref{fig:Eff} also shows the fit of an error function to the trigger efficiency curve, which is used for a verification of the trigger threshold in units of energy. The threshold is defined as the value at which 50$\,$\% of all simulated pulses that are not lost to the trigger dead time are triggered. The fit to the trigger efficiency curve with 1000 bins between 0 and 20$\,$eV results in an energy threshold of (6.70$\, \pm \,$0.22)$\,$eV and a resolution of (1.03$\, \pm \,$0.20)$\,$eV, confirming the claimed threshold.

\section{Dark matter results} \label{DMResults}
For the DM analysis we use the data set described in Sec. \ref{Analysis}. We consider all events above the energy threshold at 6.7$\,$eV up to 130$\,$eV for the DM analysis, excluding the non-linear region of the detector. All events that lie within this energy range are considered as potential DM events. We calculate our exclusion limits using the Yellin optimum interval method \cite{PhysRevD.66.032005, yellin2007extending}. We neglect the thin layer of Si on the sapphire wafer and only calculate the expected recoil spectra for Al$_{2}$O$_{3}$.

Sapphire as a detector material is suitable to search for spin-independent (see Sec. \ref{SpinInDep}) as well as spin-dependent (see Sec. \ref{SpinDep}) DM interactions. Using the simulations described in Sec. \ref{SpecAndEff}, we calculate the energy spectra we expect to measure in our detector (considering the distortions of the differential recoil spectra by the trigger efficiency, the selection criteria and the finite detector resolution) for each DM mass under the different interaction models described below. In both cases we assume the standard DM halo model, with a local DM density $\rho_{\chi} \,$=$\,$0.3$\,$GeV/c$^{2}$/cm$^{3}$ \cite{LEWIN199687} and the DM velocity following a Maxwell-Boltzmann distribution with the rotational velocity of our solar system at $v_{\odot} \,$=$\,$220$\,$km/s \cite{DONATO1998247} and the escape velocity $v_{\mathrm{esc}} \,$=$\,$544$\,$km/s \cite{10.1111/j.1365-2966.2007.11964.x}. These assumptions are in accordance with the recommended conventions in \cite{Baxter2021}.

\subsection{Spin-independent interactions} \label{SpinInDep}
\paragraph{Recoil rate:} the differential recoil rate of spin-independent elastic DM-nucleus interactions is expressed as \cite{del2022theory}:

\begin{equation} \label{eq:RecRateSI}
	\dfrac{\mathrm{d}R_{\chi}}{\mathrm{d}E_{\mathrm{R}}} = \sum_{T} f_{T} N_{T} \dfrac{\rho_{\chi}}{m_{\chi}} \dfrac{m_{T}}{2\mu_{\mathrm{N}}^{2}} A^{2} \sigma_{\mathrm{N}} F^{2}(q) \mathcal{I}(v_{\mathrm{min}}) 
\end{equation}

with $N_{T}$ being the number of target nuclei $T$ per unit mass, $m_{\chi}$ the mass of the DM particle, $m_{T}$ the mass of the target nucleus and $\mu_{\mathrm{N}}$ the DM-nucleon reduced mass. The material-independent DM-nucleon interaction cross section is $\sigma_{\mathrm{N}}$. The relation to the spin-independent zero-momentum transfer cross section on a point-like nucleus, $\sigma_{0}$, is written as \cite{del2022theory}:

\begin{equation} \label{eq:SigmaSI}
	\sigma_{0} = \sigma_{\mathrm{N}} \cdot A^{2} \cdot \dfrac{\mu_{T}^{2}}{\mu_{\mathrm{N}}^{2}}
\end{equation}

which introduces the atomic number, $A$, into eq. \ref{eq:RecRateSI}. We take the structure of the nucleus into account via the nuclear form factor $F^{2}(q)$, depending on the momentum transfer $q = \sqrt{2m_{T}E_{\mathrm{R}}}$. In this work we use the Helm form factor \cite{PhysRev.104.1466} in the parametrization of Lewin and Smith \cite{LEWIN199687}. $\mathcal{I}(v_{\mathrm{min}})$ describes the mean inverse velocity of DM particles, with $v_{\mathrm{min}}(E_{\mathrm{R}}) = \sqrt{\frac{m_{T}E_{\mathrm{R}}}{2\mu_{T}^{2}}}$, being the minimal velocity a DM particle needs to create a recoil of energy $E_{\mathrm{R}}$ in the target. An analytical description of $\mathcal{I}(v_{\mathrm{min}})$ is given in \cite{DONATO1998247}. The total recoil rate in a composite material is expressed as a sum over the different target nuclei weighted by their respective contribution to the total target mass, $f_{T}$. \\

\begin{figure}[h]
\centering
\includegraphics[width=0.48\textwidth]{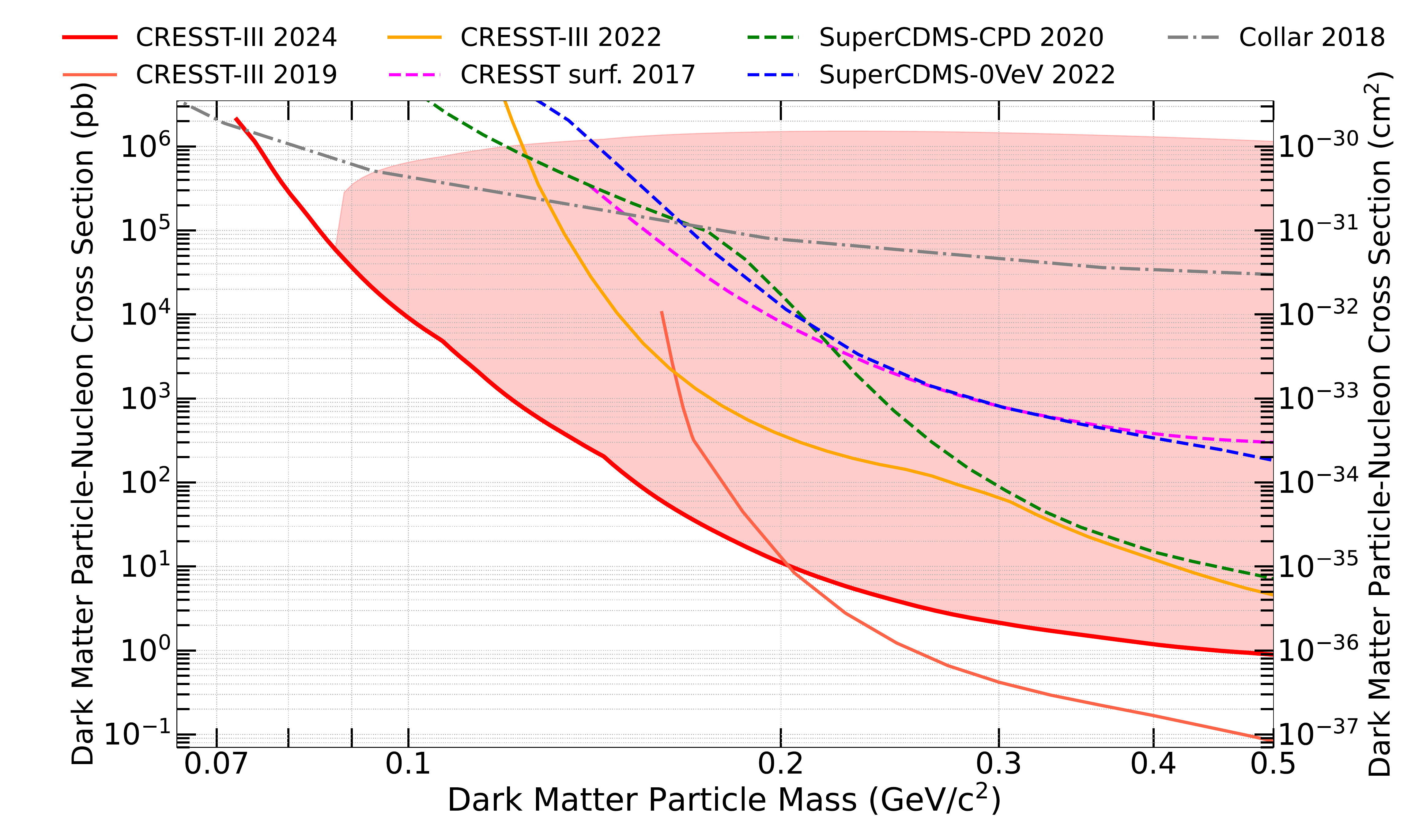}
\caption{\label{fig:SI_Limits} Upper limits on the elastic, spin-independent dark matter particle-nucleon interaction cross section as a function of the dark matter particle mass at 90$\,$\% confidence level (C.L.). The result of this work is shown in the thick solid red line with the strongest exclusion limit for dark matter masses between (74$\,$-$\,$202)$\,$MeV/c$^{2}$. Other exclusion limits obtained from solid-state cryogenic detectors are also shown, dashed lines correspond to measurements above ground: CRESST-III (Si) \cite{PhysRevD.107.122003} in orange, CRESST-III (CaWO$_{4}$) \cite{PhysRevD.100.102002} in red, CRESST-surf (Al$_{2}$O$_{3}$) \cite{Angloher2017} in magenta, SuperCDMS-CPD (Si) \cite{PhysRevLett.127.061801} in green and SuperCDMS-0VeV (Si) \cite{PhysRevD.105.112006} in blue. Shown in the dash-dotted gray line is the limit obtained from organic scintillators by J. I. Collar \cite{PhysRevD.98.023005}. The upper boundary of the shaded red area shows a very conservative estimation of the parameter space below which DM particles are not affected by the rock overburden.}
\end{figure}

\paragraph{Exclusion limits:} Figure \ref{fig:SI_Limits} shows the 90$\,$\% C.L. upper limit on spin-independent DM-nucleon interaction cross sections as a function of the DM particle mass. The limits are calculated using the Yellin optimum interval method \cite{PhysRevD.66.032005, yellin2007extending}.

Due to the extremely low threshold, the exclusion limit extends down to a DM mass of 73$\,$MeV/c$^{2}$, giving leading limits between masses of 74$\,$MeV/c$^{2}$ and 202$\,$MeV/c$^{2}$. The result improves previous limits by a factor of about 30, 400, 40 at masses of 100, 130, 165$\,$MeV/c$^{2}$, respectively. Above 202$\,$MeV/c$^{2}$ the limit is not competitive with previous limits anymore due to the high rate in the LEE, the low exposure and the low cutoff energy at 130$\,$eV. The shaded red area indicates a very conservative estimation of the parameter space above which we expect the DM particles to be affected by the rock overburden. Considering this estimate,  the exclusion limits for DM particles with masses above 87.3$\,$MeV/c$^{2}$ are not affected. Below this mass the particles are expected to undergo one or multiple scatterings in the overburden. Some details of this estimation are described in App. \ref{Overburden}.

\subsection{Spin-dependent interactions} \label{SpinDep}
\paragraph{Recoil rate:} the differential recoil rate of spin-dependent interactions of DM with protons/neutrons only, denoted as $\mathrm{p/n}$, is expressed as \cite{del2022theory}:

\begin{equation} \label{eq:RecRateSD}
	\dfrac{\mathrm{d}R_{\chi}}{\mathrm{d}E_{\mathrm{R}}} = \sum_{T} f_{T} N_{T} \dfrac{\rho_{\chi}}{m_{\chi}} 2 m_{T} \dfrac{(J_{T}+1)}{3J_{T}} \dfrac{\langle S_{\mathrm{p/n}, T} \rangle^{2}}{\mu_{\mathrm{p/n}}^{2}} \sigma_{\mathrm{p/n}} \mathcal{I}(v_{\mathrm{min}}) 
\end{equation}

with $N_{T}$, $\rho_{\chi}$, $m_{\chi}$, $m_{T}$ and $\mathcal{I}(v_{\mathrm{min}})$ having the same definition as above and $\mu_{\mathrm{p/n}}$ being the reduced mass of DM-proton/neutron. The sum runs over the isotopes of the target nuclei $T$ which are sensitive to spin-dependent interactions, weighted by their respective contribution to the total target mass $f_{T}$. We work in the limit of small momentum transfers ($q \rightarrow 0$) and therefore neglect the nuclear form factor. The relation of the material independent DM-proton/neutron cross section $\sigma_{\mathrm{p/n}}$ to the zero-momentum transfer cross section $\sigma_{0}$ is given by \cite{del2022theory}:

\begin{equation} \label{eq:SigmaSD}
	\sigma_{0} = \dfrac{4}{3} \cdot \dfrac{\mu_{T}^{2}}{\mu_{\mathrm{p/n}}^{2}} \sigma_{\mathrm{p/n}} \cdot \dfrac{J_{T}+1}{J_{T}} \cdot \langle S_{\mathrm{p/n}, T} \rangle^{2}
\end{equation}

with $J_{T}$ being the nuclear ground state angular momentum of the nucleus $T$ and $\langle S_{\mathrm{p/n}, T} \rangle$ is the expectation value of the proton/neutron spins in the nucleus $T$. For the calculation of the exclusion limits in this work we use $J = 5/2$ for both, $^{27}$Al and $^{17}$O \cite{HAASE1994171}, $\langle S_{\mathrm{p}} \rangle = 0.343$ and $\langle S_{\mathrm{n}} \rangle = 0.0296$ for $^{27}$Al \cite{PhysRevC.52.2216} and $\langle S_{\mathrm{p}} \rangle = 0$ and $\langle S_{\mathrm{n}} \rangle = 0.5$ for $^{17}$O \cite{PhysRevD.40.2131}. We use the natural abundances of 100$\,$\% for $^{27}$Al and 0.037$\,$\% for $^{17}$O \cite{HoldenCoplenBöhlkeTarboxBenefielddeLaeterMahaffyOConnorRothTepperWalczykWieserYoneda+2018+1833+2092}.
\\
\paragraph{Exclusion limits:} the 90$\,$\% C.L. upper limit on spin-dependent DM-neutron interactions is shown in \mbox{Fig. \ref{fig:SDn_Limits}}, the limit on DM-proton interactions is shown in Fig. \ref{fig:SDp_Limits}.

\afterpage{
\begin{figure}[t]
\centering
\includegraphics[width=0.48\textwidth]{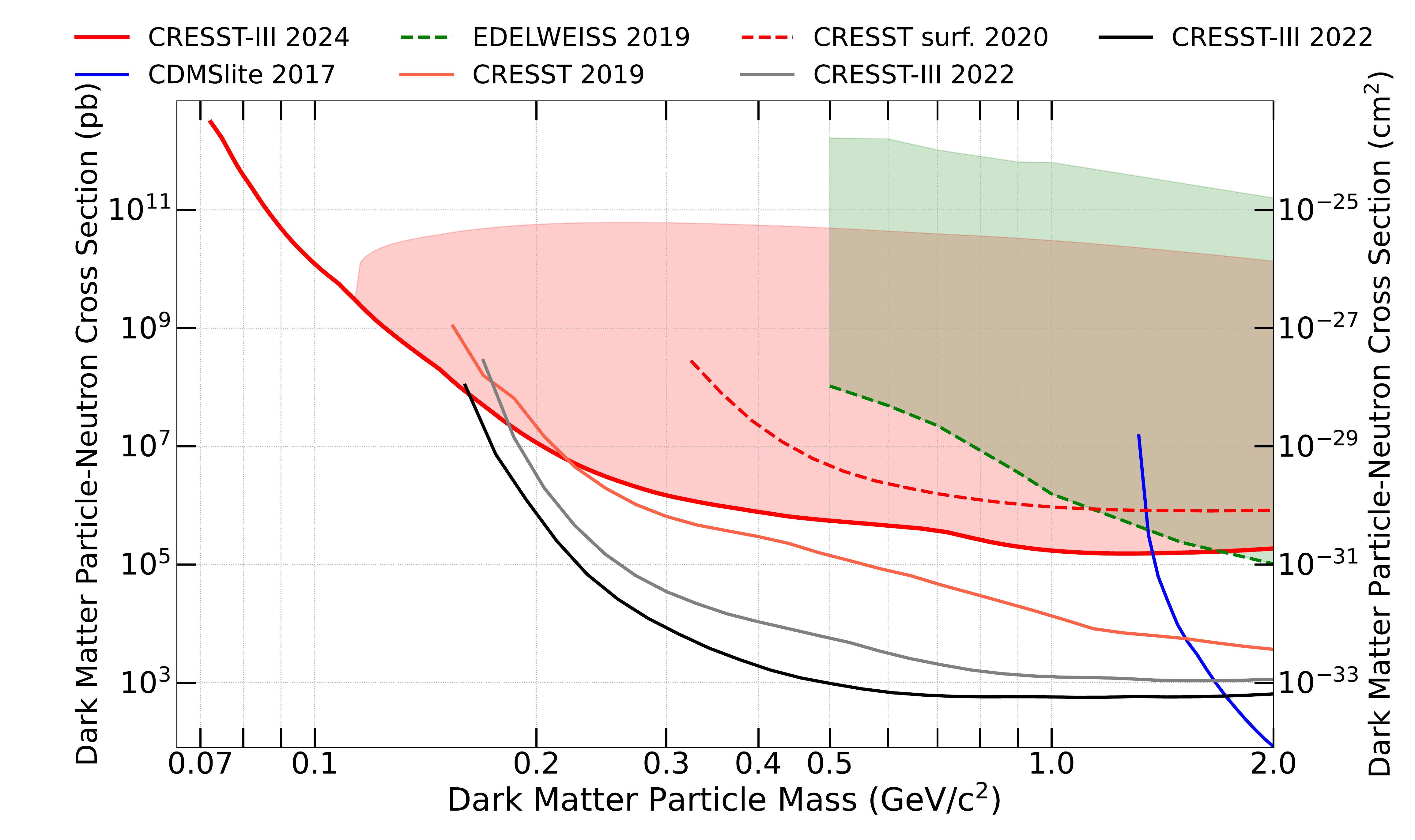}
\caption{\label{fig:SDn_Limits} Upper limits on the elastic, spin-dependent dark matter particle-neutron interaction cross section as a function of the dark matter particle mass at 90$\,$\% confidence level (C.L.). The result of this work is shown in the solid red line with the strongest exclusion limit for dark matter masses between (73$\,$-$\,$162)$\,$MeV/c$^{2}$. Other exclusion limits obtained from solid-state cryogenic detectors are also shown, dashed lines correspond to measurements above ground: EDELWEISS ($^{73}$Ge) \cite{PhysRevD.99.082003} in green (including a shaded region showing the constraints on the cross section from the overburden), CDMSlite ($^{73}$Ge) \cite{PhysRevD.97.022002} in blue, CRESST-surf (LiAlO$_{2}$) \cite{Angloher2022} in dashed red, CRESST-III (CaWO$_{4}$) \cite{PhysRevD.100.102002} in orange and the results of two CRESST-III detectors (LiAlO$_{2}$) \cite{PhysRevD.106.092008} in black and gray. The upper boundary of the shaded red area shows a very conservative estimation of the parameter space below which DM particles are not affected by the rock overburden.}
\end{figure}

\begin{figure}[t]
\centering
\includegraphics[width=0.48\textwidth]{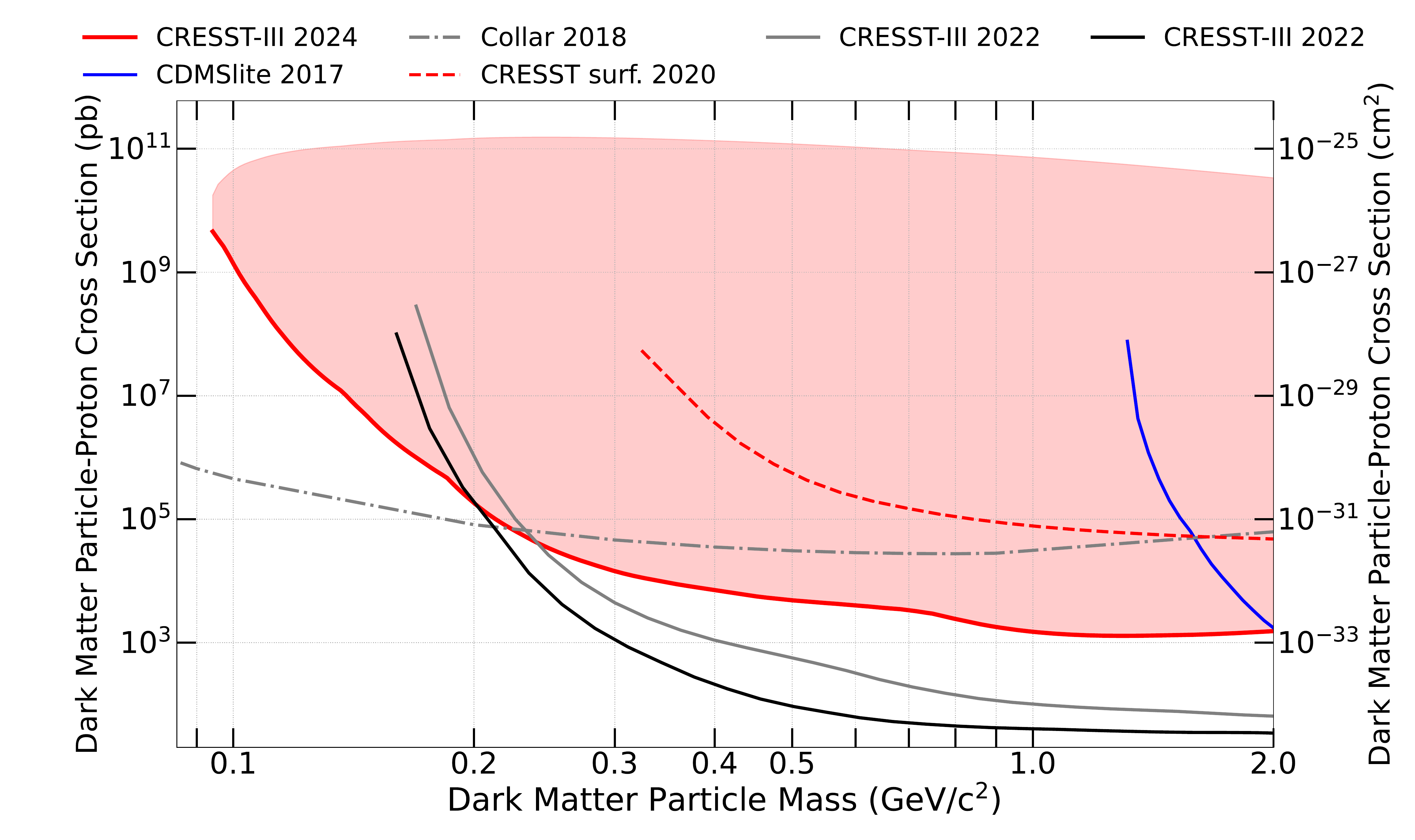}
\caption{\label{fig:SDp_Limits} Upper limits on the elastic, spin-dependent dark matter particle-proton interaction cross section as a function of the dark matter particle mass at 90$\,$\% confidence level (C.L.). The result of this work is shown in the solid red line. Other exclusion limits obtained from solid-state cryogenic detectors are also shown, dashed lines correspond to measurements above ground: CDMSlite ($^{73}$Ge) \cite{PhysRevD.97.022002} in blue, CRESST-surf (LiAlO$_{2}$) \cite{Angloher2022} in dashed red and the results of two CRESST-III detectors (LiAlO$_{2}$) \cite{PhysRevD.106.092008} in black and gray. Limits by J. I. Collar in dash-dotted gray \cite{PhysRevD.98.023005}. The upper boundary of the shaded red area shows a very conservative estimation of the parameter space below which DM particles are not affected by the rock overburden.}
\end{figure}
}

Due to the low threshold, also the limit on spin-dependent DM-neutron interactions extends down to a mass of 73$\,$MeV/c$^{2}$, giving leading limits in a mass range from 73$\,$MeV/c$^{2}$ to 162$\,$MeV/c$^{2}$. Conservatively, masses below about 113$\,$MeV/c$^{2}$ are expected to be affected by scattering in the overburden, details are given in App. \ref{Overburden}. Due to $^{17}$O not being sensitive to spin-dependent DM-proton interactions, only $^{27}$Al is considered in the calculation of the exclusion limit on DM-proton interactions. Therefore the lowest mass that can be probed is slightly higher at 94$\,$MeV/c$^{2}$. The overburden is not expected to have any influence in the entire excluded mass range in this case.

\section{Conclusion} \label{Conclusion}
In this article we report on the results of the analysis of a thin SOS wafer detector of the CRESST-III experiment. This detector has an excellent resolution of (1.0$\, \pm \,$0.2)$\,$eV and an energy threshold for nuclear recoils of (6.7$\, \pm \,$0.2)$\,$eV, allowing to probe DM particle masses below 100$\,$MeV/c$^{2}$. Furthermore, this low energy threshold made it possible to observe single photons in a CRESST detector for the first time. VUV luminescence photons with an energy of 7.6$\,$eV, emitted by a larger Al$_{2}$O$_{3}$ crystal in the vicinity of the wafer detector, irradiated by an $^{55}$Fe source, are used for a fine-tuning of the energy calibration at very low energies. The calculation of exclusion limits on spin-independent elastic DM-nucleon scattering results in leading limits in a mass range between (74$\,$-$\,$202)$\,$MeV/c$^{2}$. We expect DM particles with masses below 87.3$\,$MeV/c$^{2}$ might be affected by interactions with the overburden, this effect has to be studied in more detail. Al$_{2}$O$_{3}$ contains isotopes that are also suitable for the calculation of limits on spin-dependent interactions of DM particles on protons and neutrons. In the case of spin-dependent elastic DM-neutron scattering, the results give leading limits in a mass range between (73$\,$-$\,$162)$\,$MeV/c$^{2}$. In this case we expect masses below 113$\,$MeV/c$^{2}$ might be affected by interactions with the overburden. The presence of an excess of events at low energies, incompatible with the assumption of a flat background, strongly decreases our sensitivity in this mass regime. The origin of these events is subject to current investigations within CRESST-III and many other DM direct detection and CE$\nu$NS experiments employing solid state detectors.

\begin{acknowledgments}

We are grateful to Laboratori Nazionali del Gran Sasso - INFN for their generous support of CRESST. This work has been funded by the Deutsche Forschungsgemeinschaft (DFG, German Research Foundation) under Germany's Excellence Strategy-EXC 2094-390783311 and through the Sonderforschungsbereich (Collaborative Research Center) SFB1258 "Neutrinos and Dark Matter in Astro- and Particle Physics", by BMBF Grants No. 05A20WO1 and No. 05A20VTA and by the Austrian science fund (FWF): Grants No. I5420-N, No. W1252-N27, No. FG1, and by the Austrian research promotion agency (FFG) project No. ML4CPD. J. B. and H. K. were funded through the FWF project No. P 34778-N ELOISE. The Bratislava group acknowledges a partial support provided by the Slovak Research and Development Agency (projects No. APVV-15-0576 and No. APVV-21-0377). The computational results presented were partially obtained using the Max Planck Computing and Data Facility (MPCDF) and the CBE cluster of the Vienna BioCenter.


\end{acknowledgments}

\appendix

\section{Estimation of the parameter space affected by the overburden}\label{Overburden}

For the estimation of the effect of shielding by the overburden of the experiment we follow the method described in \cite{PhysRevD.97.115006}. We give a short outline including the numbers used for the calculation and some important points to consider when interpreting the shaded regions in the exclusion results (Fig. \ref{fig:SI_Limits} - \ref{fig:SDp_Limits}) in this work.
\\

\begin{table*}[t!]
\centering
	\caption{\label{tab:Rock}List of isotopes we consider in our calculation of the effect of the overburden on spin-dependent interactions. The natural abundances are taken from \cite{HoldenCoplenBöhlkeTarboxBenefielddeLaeterMahaffyOConnorRothTepperWalczykWieserYoneda+2018+1833+2092}. Spin expectation values from \cite{bednyakov2004nuclear} (shell model for $^{29}$Si, otherwise average values), and the values for $^{14}$N from \cite{PhysRevD.97.115006}.}
	\newcolumntype{C}{>{\centering\arraybackslash}X}
	\setlength\extrarowheight{3pt}
	\noindent
    \begin{tabularx}{\textwidth}{ C C C C C}
    \hline
    Isotope & Abundance (\%) & $J$ & $\langle S_{\mathrm{p}} \rangle$ & $\langle S_{\mathrm{n}} \rangle$ \\ \hline
    $^{13}$C & 1 & 1/2 & 0.0175 & 0.1635 \\
    $^{25}$Mg & 10 & 5/2 & 0.0565 & 0.3595 \\
    $^{39}$K & 93 & 3/2 & 0.1905 & 0.0525 \\
    $^{29}$Si & 4.67 & 1/2 & 0.016 & 0.13 \\
    $^{14}$N & 99.6 & 1 & 0.5 & 0.5 \\
    $^{15}$N & 0.4 & 1/2 & 0.136 & 0.028 \\ \hline
    \end{tabularx}
\end{table*}

\paragraph{Calculation of the affected parameter space:} The shaded regions show the parameter space in which all DM particles with the corresponding mass and cross section that are expected to be detected would have reached the detector without scattering in the overburden or atmosphere. Therefore we can safely assume that the standard description of the expected distribution of DM events holds. For cross sections above the shaded regions not all DM particles would have reached the detector without scattering. Excluding the entire parameter space where even a tiny fraction of the expected DM particles have scattered once or more in the overburden is of course a very conservative approach. Many of the scattered particles will still reach the detector, albeit with an altered velocity distribution which affects the spectral shape of the expected signal and thus the calculated limit. Methods which try to take this into account and calculate an estimate on the stopping power of the overburden often assume that the DM particles move in a straight line through the overburden while being slowed down due to scattering processes. This mostly holds for heavy DM particles, but at very low masses the particles are kinematically much more likely to have larger scattering angles, which makes it difficult to estimate the actual path length of the particle in the overburden.

The calculation of the shaded regions is dependent on the observed data. At a given DM mass the excluded cross section ($\sigma_{\mathrm{low}} \in \{\sigma_{\mathrm{nucleon,SI}}, \,\sigma_{\mathrm{neutron,SD}}, \, \sigma_{\mathrm{proton,SD}}\}$) can be directly translated into an excluded number of observed events, $N_{\mathrm{obs}}$ ($\sigma_{\mathrm{low}} \propto N_{\mathrm{obs}}$, see eq. \ref{eq:RecRateSI} and eq. \ref{eq:RecRateSD}). The expected fraction of unscattered particles at the detector, $N_{\mathrm{unsc.}}$, can be expressed as (eq. 2 in \cite{PhysRevD.97.115006}): 

\begin{equation}
    \dfrac{N_{\mathrm{unsc.}}}{N_{\mathrm{obs}}} = \dfrac{\sigma_{\chi,\mathrm{N/p/n}}}{\sigma_{\mathrm{low}}} \cdot \exp{- \sum_{A} \sigma_{\chi,A} \int dr \, n_{A}(r)}
\end{equation}

The term in the exponential is defined as the optical depth for scattering off a target nucleus $A$ over a distance $r$, with $n_{A}(r)$ being the number density and $\sigma_{\chi,A}$ the interaction cross section of DM with the nucleus, which is related to the interaction cross section of DM with a single nucleon via eq. \ref{eq:SigmaSI} for the spin-independent case and via eq. \ref{eq:SigmaSD} in the spin-dependent case. The sum runs over all relevant nuclei considered in the model of the overburden. For cross sections equal to (or smaller than) $\sigma_{\chi,\mathrm{N/p/n}}$, the number of unscattered events must be equal to (or larger than) the number of observed events. Thus, by setting the fraction $\frac{N_{\mathrm{unsc.}}}{N_{\mathrm{obs}}} = 1$, we can calculate the cross section $\sigma_{\chi,\mathrm{N/p/n}}$, giving the upper boundary of the shaded region.
\\
\paragraph{Model of the overburden:} We include the atmosphere and the rock overburden of the Laboratori Nazionali del Gran Sasso (LNGS) in our calculations. The rock consists mainly of CaCO$_{3}$ and MgCO$_{3}$ and has a density of 2.71$\, \pm \,$0.05$\,$g/cm \cite{WULANDARI2004313}. With an average depth of 1400$\,$m, this results in 3800$\,$m.w.e (meters water equivalent). The full list of all elements and their abundance in the Gran Sasso rock is: C (11.88$\,$\%), O (47.91$\,$\%), Mg (5.58$\,$\%), Al (1.03$\,$\%), Si (1.27$\,$\%), K (1.03$\,$\%), Ca (30.29$\,$\%) \cite{WULANDARI2004313}. The isotopes we consider in the case of spin-dependent interactions and their respective abundances and values for $J$, $\langle S_{\mathrm{p}} \rangle$ and $\langle S_{\mathrm{n}} \rangle$ are given in Tab. \ref{tab:Rock}. The values for $^{27}$Al and $^{17}$O are the same as in the main text. 

The atmosphere in our simple model consists of 78.084$\,$\% N$_{2}$ and 20.946$\,$\% O$_{2}$ \cite{allen2000allen, CRSHandbook}. The total mass of the atmosphere is $M = 5.15 \cdot 10^{18} \,$kg \cite{CRSHandbook2}. Assuming a uniform density of 1.2$\,$kg/m$^{3}$ \cite{Tennent1971} results in a thickness of about 8.50$\,$km. Following \cite{PhysRevD.97.123013}, we assume that DM particles arrive at the location of the detector under an angle of about 54\textdegree$\,$\cite{PhysRevD.97.115006} relative to the zenith, which is the average angle at the relevant latitude given the Earth's motion through the galactic halo. In the case of spin-dependent interactions, we consider the isotopes $^{17}$O with the same values as in the main text, $^{14}$N and $^{15}$N (see Tab. \ref{tab:Rock}).
\\
\paragraph{Interpretation:} As can be seen, the calculation of the shaded region is indirectly influenced by the presence of the LEE. If a higher (lower) number of events is observed, the interaction cross section with the overburden will be estimated at a lower (higher) value. Based on our observations in \cite{10.21468/SciPostPhysProc.12.013}, we have strong reasons to believe that the LEE is not caused by an external particle source. The LEE can be included in a likelihood based approached \cite{cresstcollaboration2024likelihood}, which mitigates its impact on the calculated limits. Since the origin of the LEE is not yet fully understood, this is only possible by modelling it with an empirical function. In this work, we choose to be conservative by calculating our limits with the Yellin optimum interval method, in which we consider all observed events as possible DM events. This means that the effect of the overburden is probably strongly overestimated in the plots shown above. Nevertheless, they give a good first estimate.


\bibliography{refs.bib}

\end{document}